\documentclass[proof]{pasj01}
\draft

\begin{document} 
\Received{}
\Accepted{}

\title{Swift/BAT and MAXI/GSC Broadband Transient Monitor}

\author{Takanori \textsc{Sakamoto}\altaffilmark{1}%
}
\altaffiltext{1}{College of Science and Engineering, Department of Physics and Mathematics, 
Aoyama Gakuin University, 5-10-1 Fuchinobe, Chuo-ku, 
Sagamihara-shi, Kanagawa 252-5258}
\email{tsakamoto@phys.aoyama.ac.jp}
\author{Ryoma \textsc{Oda}\altaffilmark{1}}
%
\author{Tatehiro \textsc{Mihara}\altaffilmark{2}}
\altaffiltext{2}{MAXI team, RIKEN, 2-1 Hirosawa, Wako, Saitama, 351-0198}
%
\author{Atsumasa \textsc{Yoshida}\altaffilmark{1}}

\author{Makoto \textsc{Arimoto}\altaffilmark{3}}
\altaffiltext{3}{Department of Physics, Tokyo Institute of Technology, 2-12-1 Ookayama, Meguro-ku, Tokyo 152-8851}
\author{Scott D. \textsc{Barthelmy}\altaffilmark{4}}
\altaffiltext{4}{NASA Goddard Space Flight Center, Greenbelt, MD 20771, USA}
\author{Nobuyuki \textsc{Kawai}\altaffilmark{3}}
\author{Hans A. \textsc{Krimm}\altaffilmark{4,5,6}}
\altaffiltext{5}{Center for Research and Exploration in Space Science and Technology (CRESST), NASA Goddard Space 
Flight Center, Greenbelt, MD 20771, USA}
\altaffiltext{6}{Universities Space Research Association, 7178 Columbia Gateway Drive, Columbia, MD 21046 USA}
\author{Satoshi \textsc{Nakahira}\altaffilmark{7}}
\altaffiltext{7}{JEM Mission Operations and Integration Center, Human Spaceflight Technology Directorate, 
Japan Aerospace Exploration Agency, 2-1-1 Sengen, Tsukuba, Ibaraki 305-8505}
\author{Motoko \textsc{Serino}\altaffilmark{2}}
%

\maketitle

\begin{abstract}
We present the newly developed broadband transient monitor using the {\it Swift} Burst Alert 
Telescope (BAT) and the {\it MAXI} Gas Slit Camera (GSC) data.  Our broadband transient 
monitor monitors high energy transient sources from 2 keV to 200 keV in seven energy 
bands by combining the BAT (15-200 keV) and the GSC (2-20 keV) data.  Currently, the daily 
and the 90-minute (one orbit) averaged light curves are available for 106 high energy transient 
sources.  Our broadband transient monitor is available to the public through our web server, 
http://yoshidalab.mydns.jp/bat\_gsc\_trans\_mon/, 
for a wider use by the community.  
We discuss the daily sensitivity of our monitor and 
possible future improvements to our pipeline.  
\end{abstract}

\section{Introduction}

High energy astrophysical sources show a temporal variability in a broad spectral range.  
The X-ray (1-10 keV) and hard X-ray ranges (10-100 keV) are especially important observing 
windows to understand the temporal characteristics of high energy sources which involve 
spectral changes.  For instance, a black hole candidate shows state changes in its 
flux and spectrum; in the quiescent state the X-ray emission is dominated by the hard emission, whereas 
in outburst the soft emission from the accretion disk becomes dominate.  
Some low mass X-ray binaries produces bright bursts in X-rays, so called X-ray bursts, which are 
caused by thermonuclear flashes of accreting material on the surface of the neutron star.  
An accreting X-ray binary pulsar sometimes shows intense outbursts in X-rays when the neutron star passes 
the disk or dense stellar wind of its companion.  Cyclotron resonance lines, one of the direct observational approaches 
to measure the magnetic field of a neutron star, are most easily observable 
in the hard X-ray range from binary pulsar during outburst.
An isolated neutron star with a high magnetic field, a so called magnetar, occasionally shows an outburst 
with multiple short duration bursts in hard X-rays 
believed originate from a large release of its internal magnetic energy.  
The time-domain astronomy which has been 
revolutionized by {\it Swift} \citep{gehrels2004} has become a frontier field of astronomy.  

The Burst Alert Telescope (BAT; \cite{barthelmy2005}) onboard {\it Swift} 
has been monitoring the hard X-ray sky (14-200 keV) thanks to its wide field of view 
since 2004.  On the other hand, the Gas Slit Camera (GSC; \cite{mihara2011}) on the {\it MAXI} 
mission \citep{matsuoka2009} has been observing the sky in softer energy band (2-30 keV) since 2009.  
Both instrument teams are providing the light curve data in real-time to the public \citep{krimm2013,sugizaki2011}.  
However, the Swift/BAT transient monitor\footnote{http://swift.gsfc.nasa.gov/results/transients/} 
\citep{krimm2013} is limited to a single energy band of 15-50 keV 
because the pipeline is using the data extracted in a single 15-50 keV 
band on-board (the data product called the BAT scaled-map data).  
Furthermore, 
there is no realtime transient monitor producing light curves in the full dynamic range of the BAT and the GSC 
data with the same time axis.  
By combining the data of the BAT and the GSC, we are able to construct broadband light curves of 
high energy transient sources with a high scientific merit.  
For example, the spectral state changes of several black hole candidates are investigated intensively 
using the BAT and the GSC data (XTE J1752$-$223, e.g., \cite{nakahira2010}; Swift J1753.5$-$0127, e.g., \cite{yoshikawa2015}; 
Swift J1910.2$-$0546, e.g., \cite{nakahira2014}; GX 339$-$4, e.g., \cite{shidatsu2011b}).  The state transition of bright low-mass 
X-ray binaries have also been studied in detail using both BAT and GSC data (e.g., \cite{asai2015}. ).  Using both the {\it MAXI} and the {\it INTEGRAL} 
data, there is the web page for monitoring 
X-ray and hard X-ray activities of high-mass X-ray binaries.\footnote{http://integral.esac.esa.int/bexrbmonitor/webpage\_oneplot.php}
Therefore there is a great demand to gather broad band light curves of high-energy transient sources from multiple missions and 
present them in a consistent format along a single time axis.  

In this paper, we 
introduce the broadband transient monitor utilizing the BAT and GSC data.  This transient monitor 
covers the dynamic range from 2 keV to 200 keV, which is ideal to monitor high energy 
transients in a broad spectral coverage.  

The paper is organized as follows.  We present the analysis method in \S 2.  In \S 3, our broadband 
transient monitor is introduced.  We further discuss our broadband transient monitor in \S 4.  

\section{Analysis}

\begin{figure}
 \begin{center}
  \includegraphics[width=12cm]{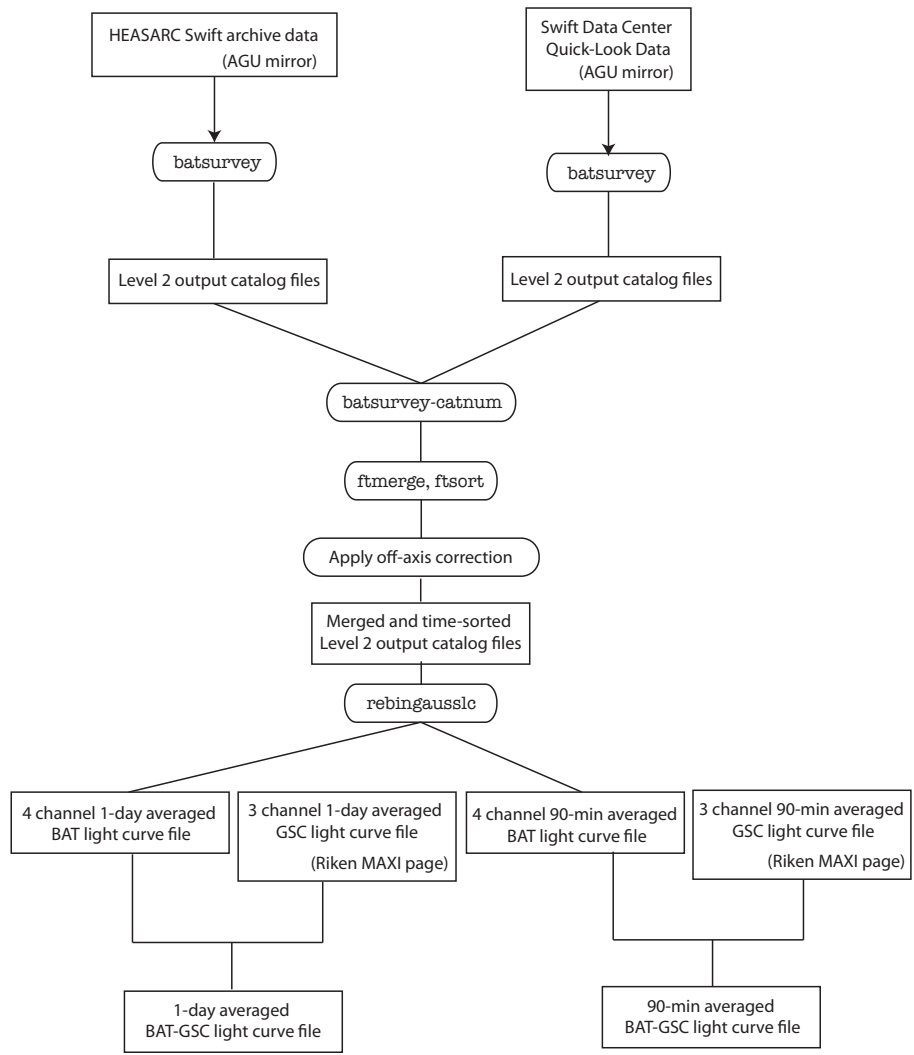} 
 \end{center}
\caption{Flowchart of the BAT-GSC transient monitor pipeline.}\label{fig:flowchart}
\end{figure}

The flowchart of our pipeline is shown in Figure \ref{fig:flowchart}.  
As the initial step to construct the broadband transient monitor, 
independent pipelines were developed just 
to mirror the {\it Swift} BAT data to our server.  This step is rather crucial for the entire process 
because it takes a significant amount of time to download the BAT data from the {\it Swift} data 
archive center in the U.S. (Swift Data Center (SDC) \footnote{http://swift.gsfc.nasa.gov/cgi-bin/sdc/ql?} and HEASARC\footnote{http://heasarc.gsfc.nasa.gov/FTP/swift/data/obs/}) to Japan.  
Since the {\it Swift} data are initially stored at the SDC, and then, archived to the HEASARC 
seven days after the observation, two separate pipelines were developed to mirror both SDC and HEASARC data,   
which are available to the public from our web server 
(the SDC data mirror\footnote{http://yoshidalab.mydns.jp/swift\_sdc\_ql/} and the HEASARC data 
mirror\footnote{http://yoshidalab.mydns.jp/swift\_bat\_heasarc/}) located in Japan.  
Note that only the Swift/BAT related data are mirrored from the HEASARC to our server.
The script to mirror the SDC data is running once an hour, whereas the script to mirror 
the HEASARC archive runs once a day. 

The basic BAT data analysis is performed using the HEASOFT software package.  
The {\tt batsurvey} script is used to process the BAT survey (Detector Plane 
Histogram (DPH)) data.  The default eight energy bands (14-20 keV, 20-24 keV, 
24-35 keV, 35-50 keV, 50-75 keV, 75-100 keV, 100-150 keV and 150-195 keV) and 
the original time resolution of the DPH data (timesep = "DPH") were specified in the 
script.  The typical exposure time of the original DPH data is five minutes.  
Products are made for 
146 sources which were flagged as bright hard X-ray 
sources in the BAT 70 month all-sky hard X-ray survey \citep{baumgartner2013}.  
The data after August 2009, when {\it MAXI} and {\it Swift} operations overlap, 
were processed.  
The {\tt batsurvey} script produces so-called `level 2' catalogs for each 
source, which contain the extracted count-rate and the incident angle of the source.  
Once the {\tt batsurvey} process is completed, all the level 2 
catalogs (final outputs of the {\tt batsurvey} script) are merged using {\tt batsurvey-catnum} 
script and time sorted by {\tt ftsort} for all 146 sources.  

\begin{figure}
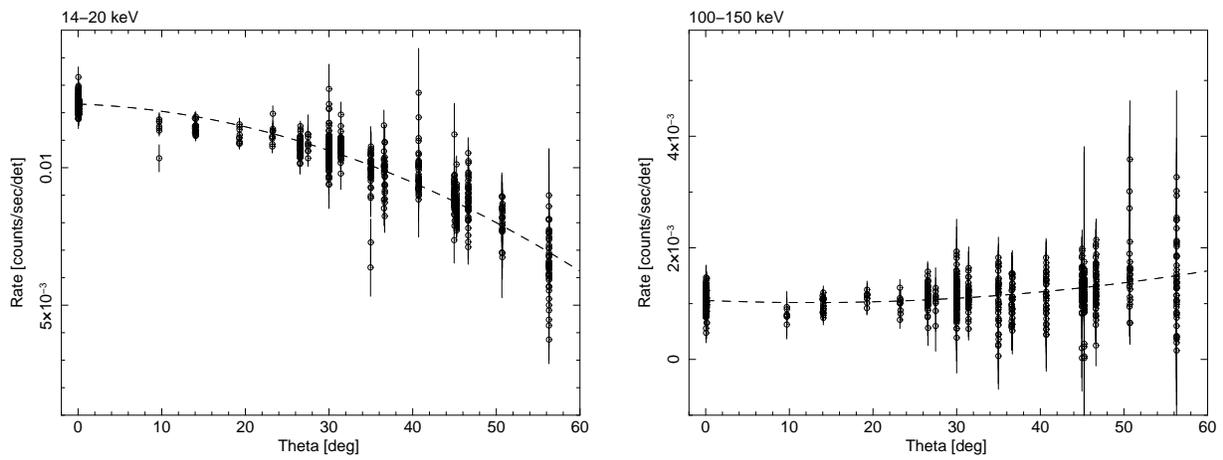

\begin{center}
\includegraphics[width=6cm,angle=-90]{crab_14_25kev_corr.ps}
\includegraphics[width=6cm,angle=-90]{crab_100_150kev_corr.ps}
\end{center}
\caption{The systematic trends in the BAT observed count rate of the Crab nebula 
as a function of the incident angle ($\theta$) in the 14-20 keV band (left) and the 
100-150 keV band (right).  The dashed line is the best fit quadratic function (table \ref{tbl:offaxis_corr}) 
to the data.}\label{fig:crab_offaxis}
\end{figure}

The BAT count rates extracted by the {\tt batsurvey} script must be corrected for the 
energy dependent vignetting of incoming photons \citep{tueller2010}.  
To model this off-axis effect in the count rate at each energy band, 
we processed the survey data between 2004 and 2005, when 
the Crab nebula was in the field of view at various incident angles.  
Figure \ref{fig:crab_offaxis} illustrates 
the significant energy dependent systematic effect between the Crab count rate and incident 
angle in the 15-20 keV and the 100-150 keV band, as an example.  
A quadratic function was used to fit 
the trend between the count rate and the incident angle for all eight energy bands (Table \ref{tbl:offaxis_corr}).  
%
In each band, the count rate is corrected by the rate of the on-axis Crab rate to the estimated Crab 
rate at the given incident angle.  
Next, the original eight energy bands are binned 
to four energy bands (14-24 keV, 24-50 keV, 50-100 keV and 100-195 keV).  And then, a one day 
averaged and a 90-minutes (1 orbit) averaged light curves are generated using {\tt rebingausslc}.  

The {\it MAXI} GSC one day and 90 minutes light curves are downloaded from the {\it MAXI} public web 
page\footnote{www.maxi.riken.jp} for the common sources of the BAT 146 bright sources in the input 
catalog of the BAT data process and the 369 (as of July 2015) sources listed in the {\it MAXI} public web page.  The 
number of common sources in the current monitor is 106.  This limitation mainly comes from the 
number of sources in the BAT input catalog.  However, since all the created BAT sky images are 
stored in our computer, only the source extraction tool, {\tt batcelldetect}, is needed to run through 
all the archival images to add a new source from the BAT data.  
We have a plan 
to add the sources which were detected in outburst in the BAT transient monitor \citep{krimm2013} in the past six years 
to our transient monitor pipeline.

The light curves of the BAT and the GSC are 
combined and plots are generated using the python matplotlib module.\footnote{http://matplotlib.org} 
The BAT light curve data are available in FITS format.  The interactive light curve based on 
the python mpld3 module\footnote{http://mpld3.github.io} is also 
available, so that a user can move and zoom-in the light curve interactively.  

\section{BAT and GSC Broadband Transient Monitor}

Our broadband transient monitor is available to the public from the web server 
at Aoyama Gakuin University: http://yoshidalab.mydns.jp/bat\_gsc\_trans\_mon/.  
Currently, 
the broadband light curves of 106 known sources are accessible from the web 
page.  The web page updates $\sim$3 times a day depending on the amount of the data 
which need to be processed.  The products of our BAT-GSC 
broadband transient monitor are summarized in Table \ref{tbl:bb_product_summary}.  

\begin{figure}
\begin{center}
\includegraphics[width=12cm]{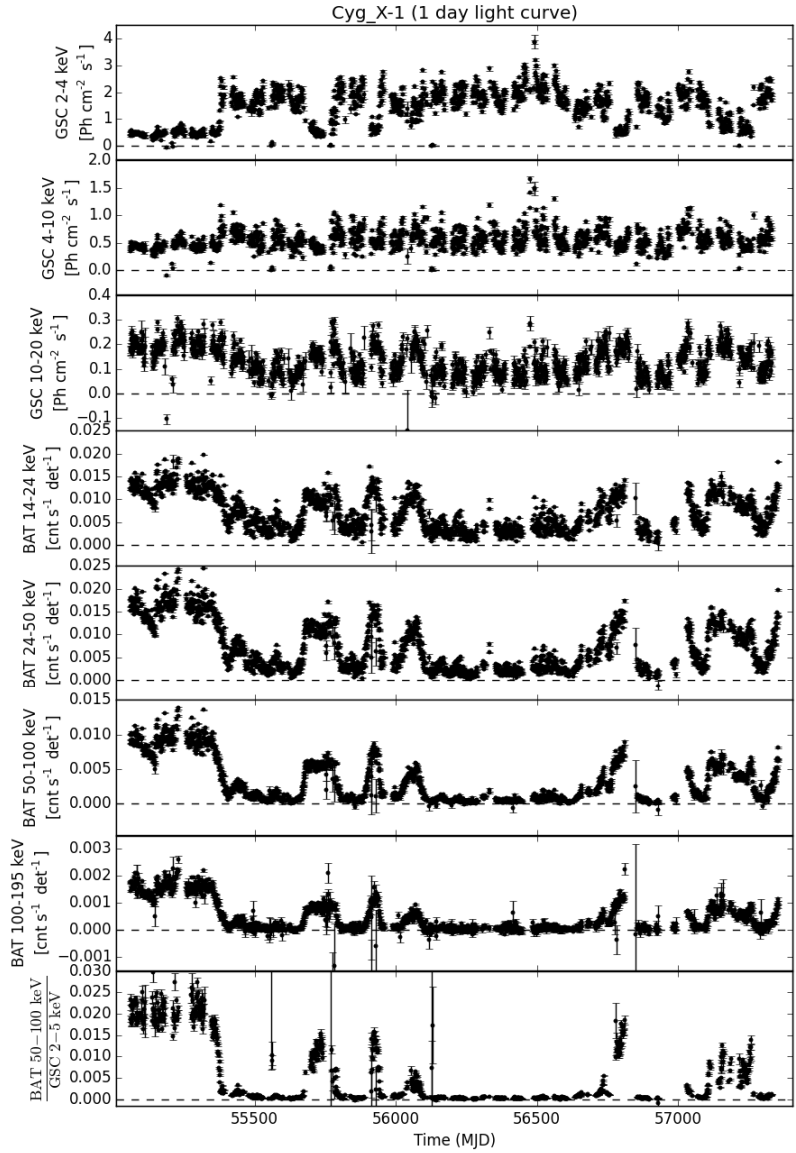}
\end{center}
\caption{Daily averaged broadband light curve of Cyg X-1.  
A clear hard-to-soft state transition is visible at MJD 55300 
and also the short soft-to-hard and hard-to-soft state transitions are visible at 
MJD 55700, 55900 and 56800.  
}\label{fig:bb_lc_cygx1}
\end{figure}

\begin{figure}
\begin{center}
\includegraphics[width=12cm]{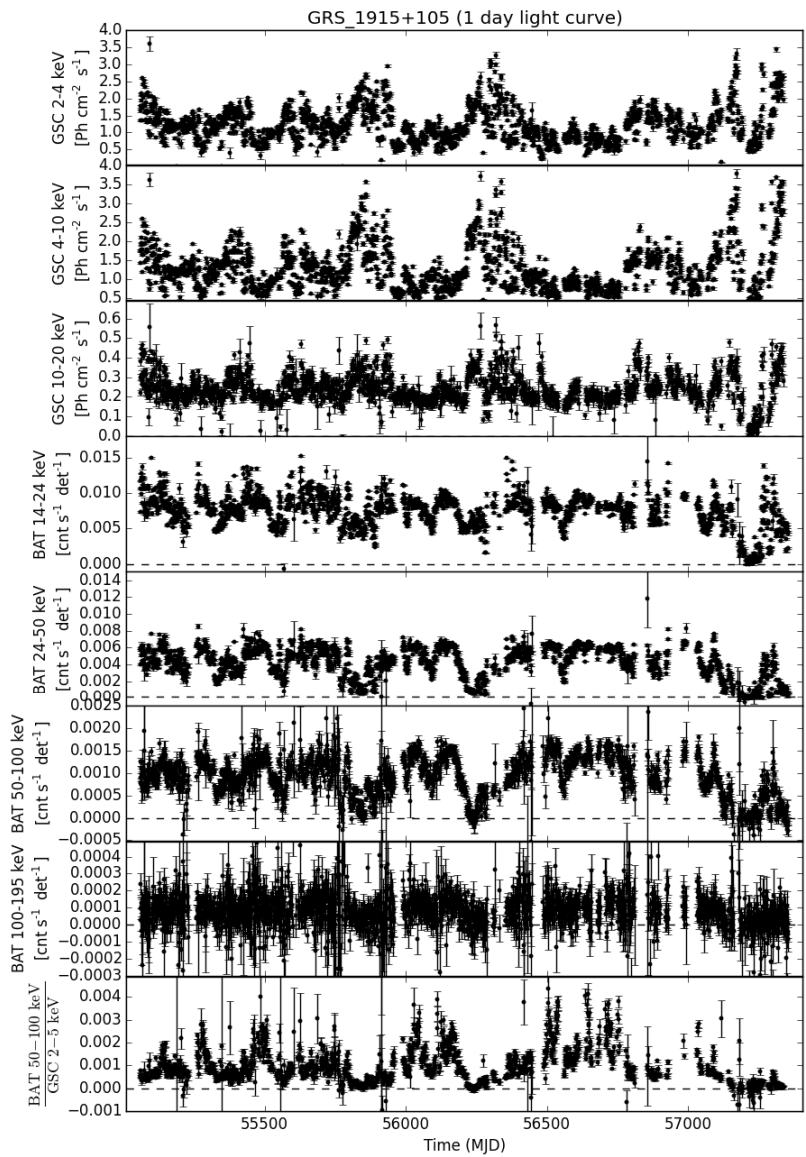}
\end{center}
\caption{Daily averaged broadband light curve of GRS 1915+105.  
Several hard-to-soft state transitions are clearly visible 
in MJD 55700, 56300 and 57200.  
}\label{fig:bb_lc_grs1915}
\end{figure}

\begin{figure}
\begin{center}
\includegraphics[width=12cm]{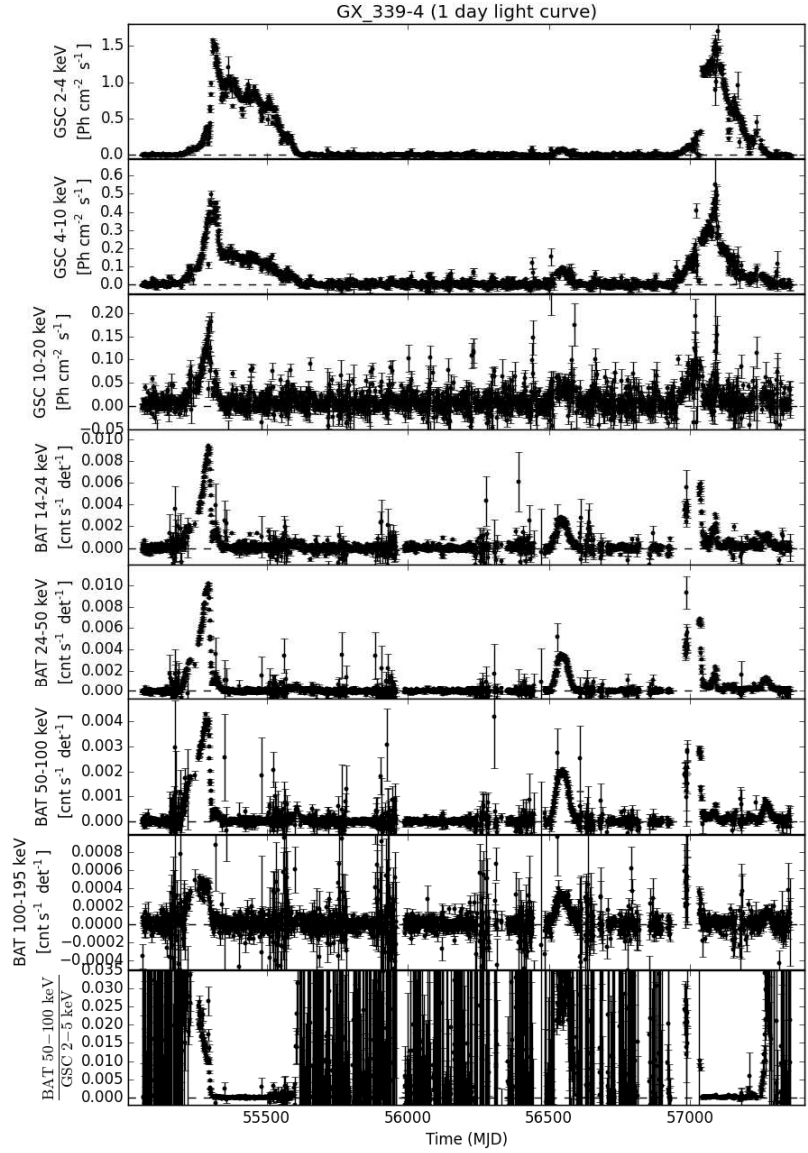}
\end{center}
\caption{Daily averaged broadband light curve of GX 339-4.
During the outburst around MJD 55200, the hard-to-soft 
spectral evolution at the tail part of its outburst is evident between 
the GSC 2-4 keV and BAT 50-100 keV band.  
}\label{fig:bb_lc_gx339}
\end{figure}

Here, we highlight the products of several sources in our broadband transient monitor.  
Figure \ref{fig:bb_lc_cygx1} and \ref{fig:bb_lc_grs1915} show our broadband light curves  
of the black hole binaries Cygnus X-1 and GRS 1915+105.  Those light curves clearly show 
multiple spectral state transitions between ``low-hard" and ``high-soft" states, 
which are believed to indicate a change in the geometry of the accretion disk (e.g., \cite{esin1997}).  
For example, the last clear state change was happening around MJD 57125 for Cygnus X-1 and 
around MDJ 56252 for GRS 1915+105.  
As can be seen in the light curves, the borderline energy which the emission becomes brighter or dimmer 
when the source is in the high-soft state is around 10 keV for Cygnus X-1 and 15 keV for GRB 1915+105.  
Figure \ref{fig:bb_lc_gx339} shows the long-term temporal behavior of GX 339-4, a Galactic transit black hole with a 
low-mass companion.  During the giant outburst in the year 2010 (e.g., MJD 55200-55600), 
the burst emission showed a strong hard-to-soft evolution.  The initial hard emission was visible up to the 
highest energy band of 100-195 keV.  This hard X-ray emission episode 
is dominated by Comptonized photons from the accretion disk (e.g., \cite{shidatsu2011a}).
%

\begin{figure}
\begin{center}
\includegraphics[width=12cm]{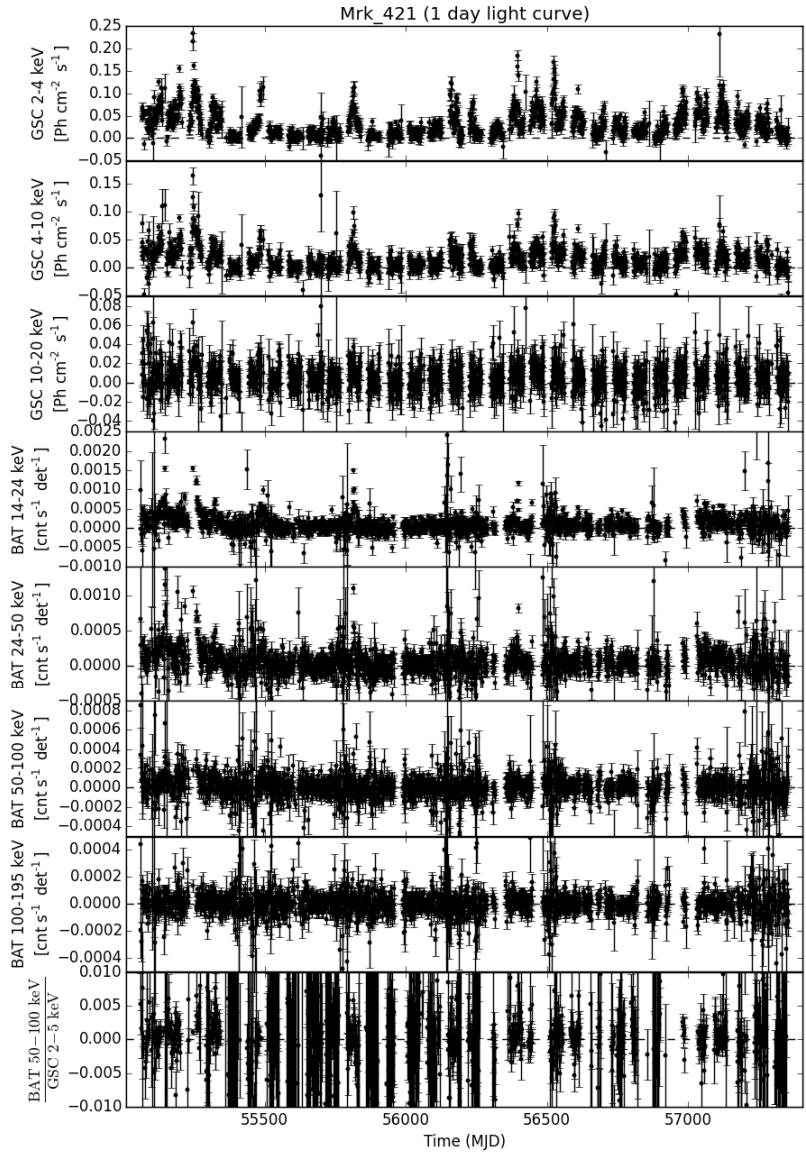}
\end{center}
\caption{Daily averaged broadband light curve of Mrk 421.  
The hard X-ray emission is detected by the BAT data 
during the outburst episodes such as MJD 55250 and 55800.  
}\label{fig:bb_lc_mrk421}
\end{figure}

\begin{figure}
\begin{center}
\includegraphics[width=12cm]{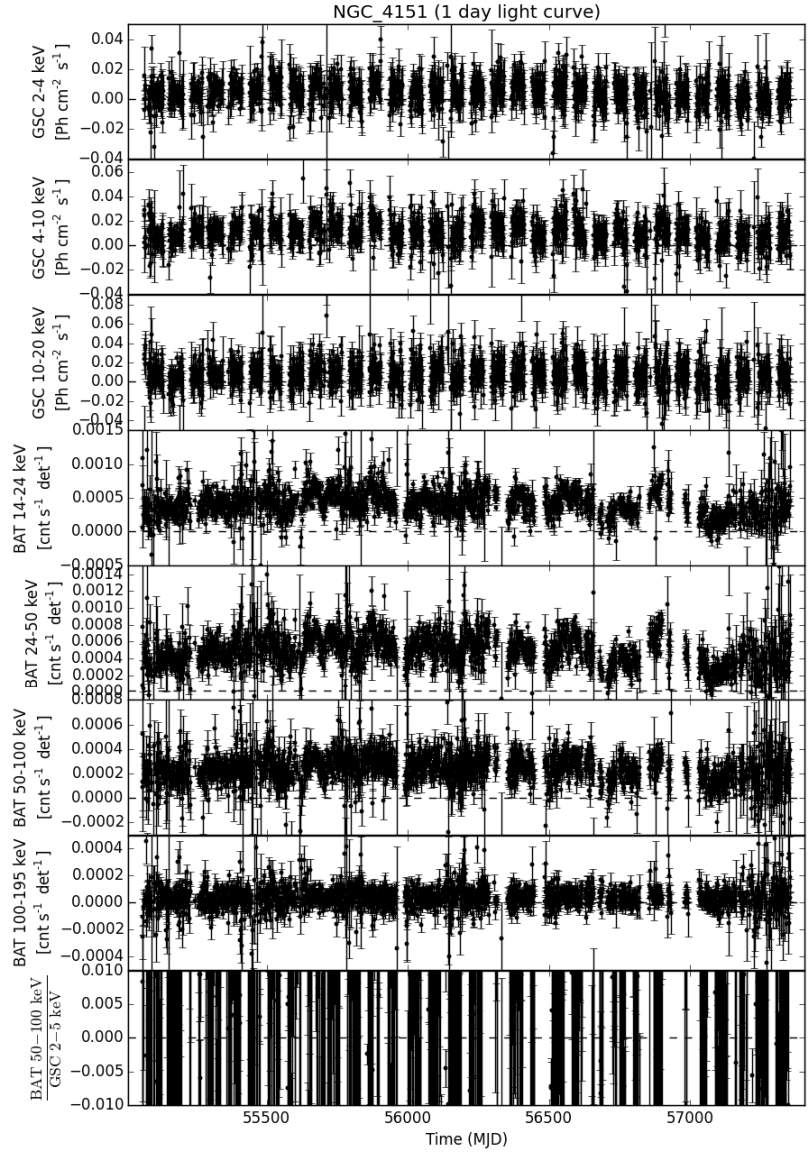}
\end{center}
\caption{Daily averaged broadband light curve of NGC 4151.  
The variable emission from NGC 4151 is clearly visible in the hard X-ray bands 
rather than the soft X-ray bands.  
}\label{fig:bb_lc_ngc4151}
\end{figure}

The $\gamma$-ray blazar, Mrk 421, showed several flares visible 
up to 100 keV (Figure \ref{fig:bb_lc_mrk421}).  
Since the BAT energy range is located at the dip 
between the two broad peaks in the spectral energy distribution of Mrk 421 (e.g., \cite{abdo2011}), 
the BAT hard X-ray emission is not that evident compared to the soft X-ray band of the GSC.  However, 
during the outburst, the hard X-ray emission is clearly visible in the BAT data, and 
the broad-spectral properties of the source can be investigated using our monitor.  
Figure \ref{fig:bb_lc_ngc4151} shows 
the light curve of the Seyfert 1.5 galaxy NGC 4151 which has a hard continuum in its spectrum 
\citep{keck2015}.  Unlike previous examples, its emission and 
the temporal variability are clearly visible in the BAT data rather than the GSC data.  

\begin{figure}
\begin{center}
\includegraphics[width=8cm]{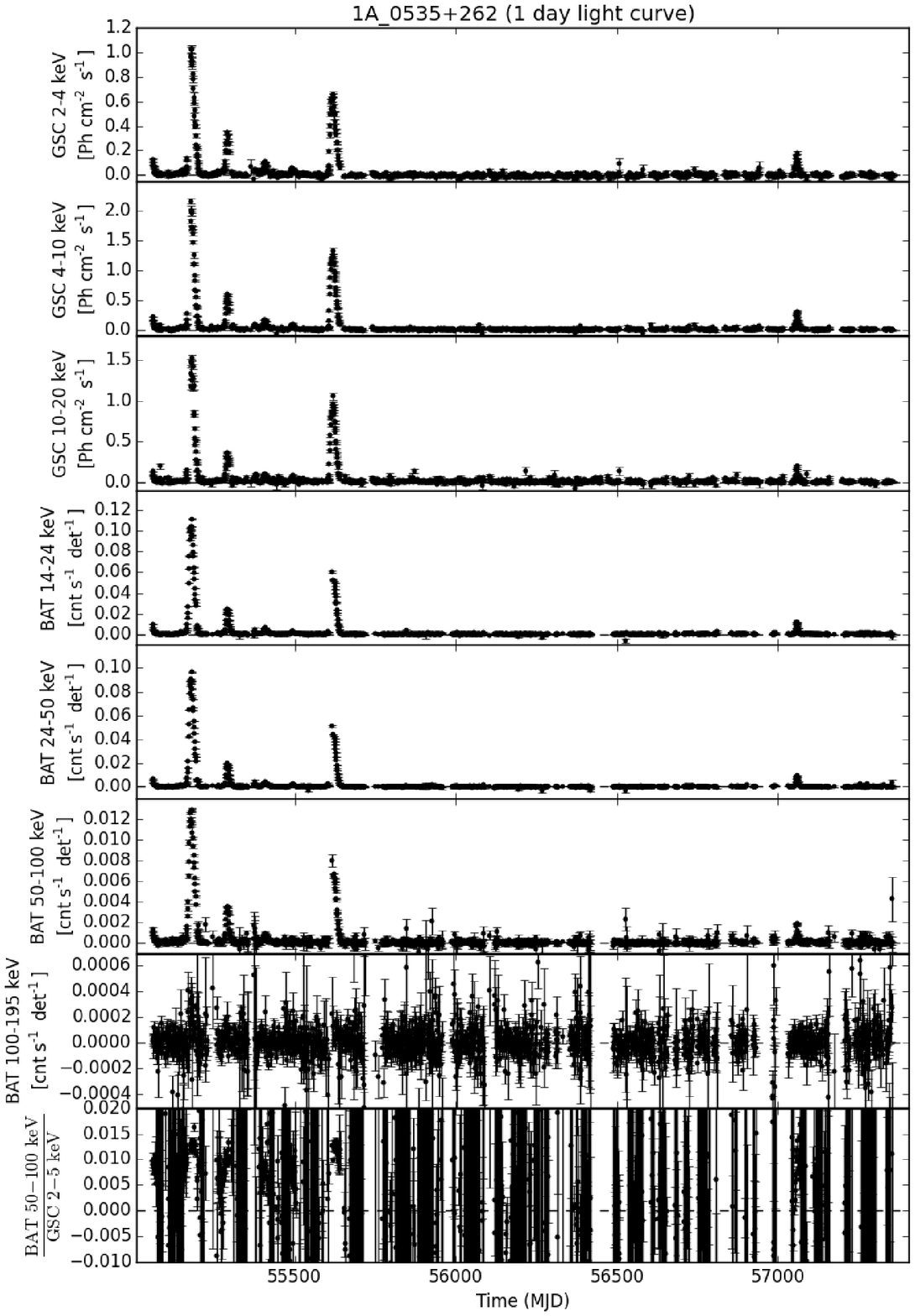}
\includegraphics[width=8cm]{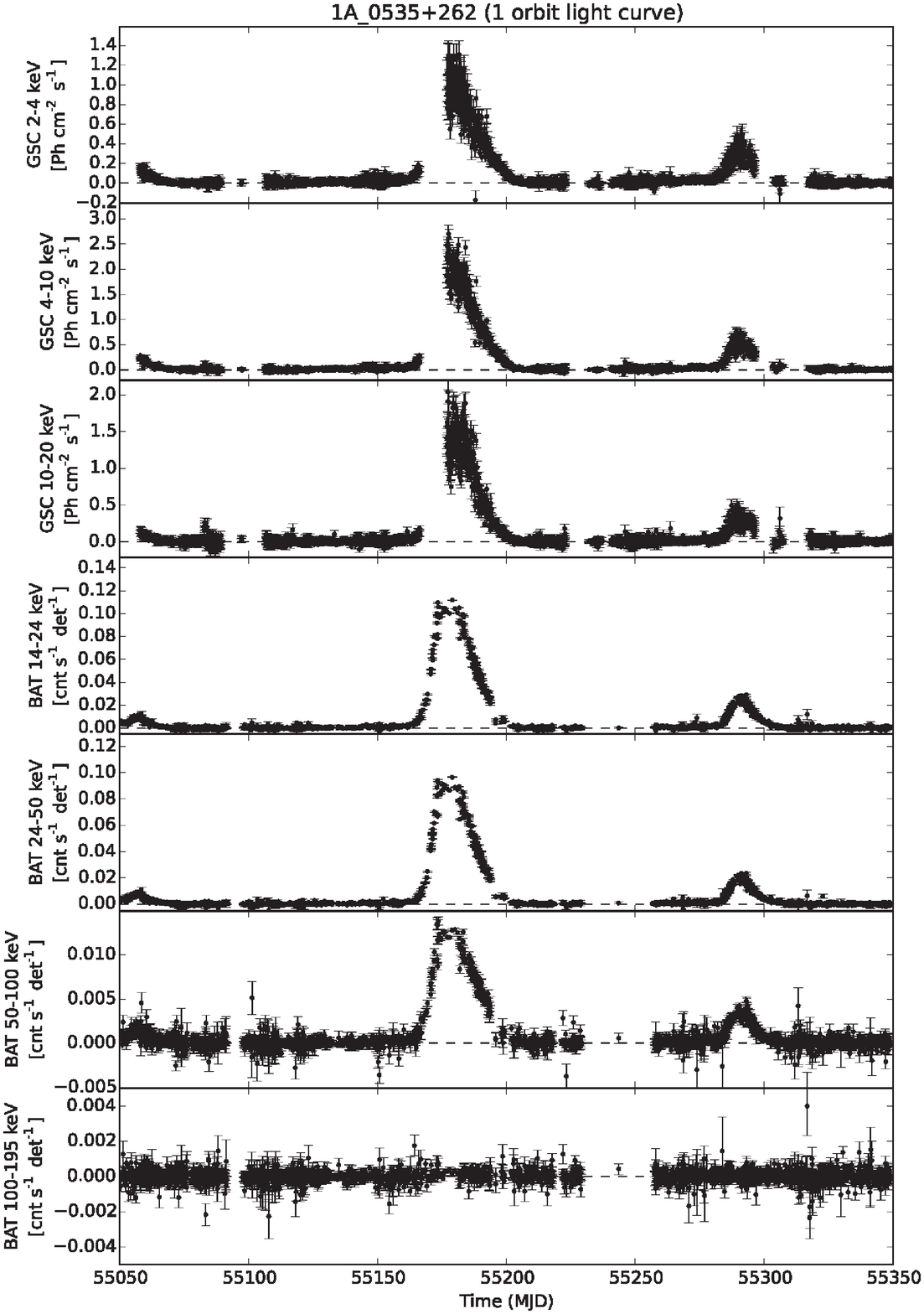}
\end{center}
\caption{Daily (left) and 90-minutes averaged (right) broadband light curve of AO535+262.  
The outbursts are clearly visible up to 100 keV.  Moreover, since the outburst 
emission is so intense, the temporal profile is clearly resolved in 90-minutes averaged light curve.  
}\label{fig:bb_lc_a0535}
\end{figure}

In Figure \ref{fig:bb_lc_a0535}, the several giant outburst 
episodes are visible in the broadband light curve of a high-mass X-ray pulsar A0535+262.  
Since the emission during the giant outbursts were so intense, 
the temporal profile up to 100 keV is clearly visible even in the one orbit light curve (right panel of 
Figure \ref{fig:bb_lc_a0535}).  A recent study showed that the energy of the cyclotron resonance 
line of A0535+262, which was found typically around 45 keV, increased when the flux was high \citep{sartore2015}.  
Thus, it is important to monitor those cyclotron sources in a broad energy band to understand the 
transient spectral features in various flux levels.  As can be seen in those examples, our transient 
monitor includes various types of sources and can monitor interesting temporal and spectral stages of high energy sources.

\section{Future Prospects}

We constructed the broadband transient monitor using the {\it Swift} BAT and the {\it MAXI} GSC 
data and made it available to the public.  This work is the first attempt to analyse the BAT survey 
data in real-time for constructing multi-band light curves.  Although the updating frequency of the 
light curve is still not ideal, we can now run the pipeline several times a day in real-time.  

Although it varies based on the {\it Swift} pointing, a typical exposure time of an individual object for the BAT 
is roughly four hours per day (about eight hours at the highest day).  Applying the BAT survey sensitivity 
equation described in \citet{baumgartner2013}, this typical exposure corresponds to a 5-sigma sensitivity of 
$\sim$10 mCrab (14-195 keV).  On the other hand, the daily averaged 5-sigma sensitivity of the GSC 
is $\sim$15 mCrab in the 4-10 keV band \citep{sugizaki2011}.  Therefore, the daily sensitivity of the BAT and the GSC are comparable.  
Therefore, it is a good match to combine the BAT and the GSC data to monitor high energy transient sources 
on a daily basis. 

All the {\it MAXI} GSC light curve data in the current broadband transient monitor is downloaded 
from the {\it MAXI}'s public web page.  Since all of the light curves are created using an aperture 
photometry method \citep{sugizaki2011}, the count rates in the light curves could be contaminated with bright X-ray sources 
located near the object.  Due to a relatively poor position resolution of the GSC, if other bright X-ray 
source is located within 2 degrees from the object, there might be an issue in the light curve.  
For those sources which could be affected by nearby contaminations, we 
have a future plan to generate the GSC light curves of the source using the point spread function 
(PSF) fit method which should provide a contamination free light curve \citep{hiroi2011,morii2015} using the 
GSC data.  At the current stage, the sources which are located within 2 degrees from 
catalog sources in the ROSAT bright source catalog \citep{voges1999} with the count rate greater than 
1 count s$^{-1}$ are flagged as possible 
contaminated sources at the top page of the monitor (see Table \ref{tbl:src_list}).  

The processing time of the current pipeline is limited by the analysis of the BAT survey data.  
Our pipeline is processing the BAT survey data at the finest time resolution with the eight energy 
bands.  Although we can process the data in the coarser time resolution and the smaller energy 
bands to speed up the process, the best spectral and temporal information can be extracted in the current 
setup for the BAT data.  For example, the outputs of our pipeline can provide 8 channel 
spectral data of the sources every five minutes when high energy sources are in the very 
bright state.  We have a plan to process the data with multiple available computers or 
Graphical Processing Unit (GPU) to speed up the process.  

The current broadband transient monitor contains 106 sources.  However, it is not a difficult 
task to add sources to the monitor.  We are planning to add sources based on not only the information 
from various observatories (e.g., the Astronomer's Telegram\footnote{http://www.astronomerstelegram.org}, 
the Gamma-ray Coordinates Network\footnote{http://gcn.gsfc.nasa.gov/gcn\_main.html}) but also 
requests from the community.  

\begin{ack}
We would like to thank the anonymous referee for comments and suggestions that materially 
improved the paper.  
This work was supported by JSPS Grant-in-Aid for Scientific Research (C) Grant Number 25400234.  
\end{ack}

\begin{longtable}{*{4}{c}}
\caption{Ths source list of the broadband transient monitor}
\endfirsthead
\caption[]{Ths source list of the broadband transient monotor\\}\label{tbl:src_list}
\endhead
\hline
Name & R.A. & Dec. & Possible contaminated source\\
\hline
V709 Cas & 7.2036 & 59.2894 & \\
Mrk 348 & 12.1964 & 31.9570 & \\
Gamma Cas & 14.1772 & 60.7167 &\\
SMC X-1 & 19.2714 & -73.4433 & T \\
2S 0114+650 & 19.5112 & 65.2916 &\\
4U 0115+634 & 19.6330 & 63.7400 &\\
CC Eri & 38.5940 & -43.7960 &\\
GK Per & 52.7993 & 43.9047 &\\
BQ Cam & 53.7495 & 53.1732 &\\
X Per & 58.8462 & 31.0458 &\\
LSV+44 17 & 70.2470 & 44.5300 &\\
LMC X-4 & 83.2075 & -66.3705 & T\\
Crab Nebula Pulsar & 83.6332 & 22.0145 &\\
1A 0535+262 & 84.7274 & 26.3158 &\\
LMC X-3 & 84.7342 & -64.0823 & T\\
NGC 2110 & 88.0474 & -7.4562 &\\
4U 0614+091 & 94.2804 & 9.1369 &\\
MXB 0656-072 & 104.5703 & -7.2105 &\\
EXO 0748-676 & 117.1388 & -67.7500 &\\
Vela Pulsar & 128.8361 & -45.1764 &\\
GS 0834-430 & 128.9790 & -43.1850 &\\ 
Vela X-1 & 135.5286 & -40.5547 &\\
MCG -05-23-016 & 146.9173 & -30.9489 &\\
GRO J1008-57 & 152.4417 & -58.2917 &\\
NGC 3227 & 155.8774 & 19.8651 & T\\
Mrk 421 & 166.1138 & 38.2088 &\\
NGC 3516 & 166.6979 & 72.5686 &\\
XTE J1118+480 & 169.5450 & 48.0370 &\\
Cen X-3 & 170.3158 & -60.6230 & T\\
NGC 3783 & 174.7572 & -37.7386 & \\
1E 1145.1-6141 & 176.8692 & -61.9539 &\\
NGC 4151 & 182.6357 & 39.4057 &\\
GX 301-2 & 186.6567 & -62.7706 &\\
3C 273 & 187.2779 & 2.0524 &\\
GX 304-1 & 195.3217 & -61.6019 &\\
Cen A & 201.3651 & -43.0191 &\\
4U 1323-619 & 201.6504 & -62.1361 &\\
IC 4329A & 207.3303 & -30.3094 &\\
NGC 5506 & 213.3119 & -3.2075 &\\
PSR B1509-58 & 228.4813 & -59.1358 &\\
Cir X-1 & 230.1703 & -57.1667 &\\
H 1538-522 & 235.5971 & -52.3861 &\\
1E 1547.0-5408 & 237.7255 & -54.3066 &\\
H 1553-542 & 239.4512 & -54.4150 &\\
4U 1608-522 & 243.1792 & -52.4231 & T \\
Sco X-1 & 244.9795 & -15.6402 &\\
SWIFT J1626.6-5156 & 246.6510 & -51.9428 & T\\
4U 1624-490 & 247.0118 & -49.1985 & \\
4U 1626-67 & 248.070 & -67.4619 &\\
4U 1636-536 & 250.2313 & -53.7514 &\\
GX 340+0 & 251.4488 & -45.6111 &\\
GRO J1655-40 & 253.5006 & -39.8458 & T\\
Her X-1 & 254.4576 & 35.3424 & T\\
OAO 1657-415 & 255.1996 & -41.6731 & T\\
XTE J1701-407 & 255.4349 & -40.8583 & T\\
GX 339-4 & 255.7063 & -48.7897 &\\
GX 349+2 & 256.4354 & -36.4231 & T\\
4U 1705-440 & 257.2270 & -44.1020 & T\\
4U 1722-30 & 261.8883 & -30.8019 & T\\
GX 9+9 & 262.9342 & -16.9617 &\\
4U 1728-34 & 262.9892 & -33.8347 &\\
GX 1+4 & 263.0090 & -24.7456 & T\\
MXB 1730-335 & 263.3504 & -33.3877 &\\
SLX 1735-269 & 264.5667 & -27.0044 & T\\
4U 1735-44 & 264.7429 & -44.4500 &\\
SGR A Gal center complex & 266.4168 & -29.0078 &\\
2E 1742.9-2929 & 266.5229 & -29.5153 &\\
XTE J17464-3213 & 266.5650 & -32.2335 & T\\ 
GX 3+1 & 266.9833 & -26.5636 & \\
SWIFT J1753.5-0127 & 268.3679 & -1.4525 &\\
GX 5-1 & 270.2842 & -25.0792 & T\\
GX 9+1 & 270.3846 & -20.5289 &\\
SAX J1808.4-3658 & 272.1150 & -36.9790 &\\
XTE J1810-189 & 272.6079 & -19.0700 &\\
SAX J1810.8-2609 & 272.6850 & -26.1500 &\\
GX 13+1 & 273.6315 & -17.1574 & \\
GX 17+2 & 274.0058 & -14.0364 & T\\
4U 1820-30 & 275.9186 & -30.3611 &\\
4U 1822-371 & 276.4450 & -37.1053 &\\
Ginga 1826-24 & 277.3675 & -23.7969 &\\
XB 1832-330 & 278.9338 & -32.9914 &\\
Ser X-1 & 279.9896 & 5.0358 &\\
Ginga 1843+00 & 281.4125 & 0.8917 &\\
4U 1850-086 & 283.2703 & -8.7057 &\\
V1223 Sgr & 283.7593 & -31.1635 &\\
XTE J1855-026 & 283.8804 & -2.6067 &\\
XTE J1856+053 & 284.1786 & 5.3094 &\\
HETE J1900.1-2455 & 285.0360 & -24.9205 &\\
SWIFT J1910.2-0546 & 287.5950 & -5.7990 &\\
Aql X-1 & 287.8167 & 0.5850 &\\
SS 433 & 287.9565 & 4.9827 &\\
GRS 1915+105 & 288.7983 & 10.9456 &\\
4U 1916-053 & 289.6995 & -5.2381 &\\
SWIFT J1922.7-1716 & 290.6542 & -17.2842 &\\
4U 1954+31 & 298.9264 & 32.0970 &\\
Cyg X-1 & 299.5903 & 35.2016 &\\
Cygnus A & 299.8682 & 40.7339 &\\
V404 Cyg & 306.0159 & 33.8672 &\\
EXO 2030+375 & 308.0633 & 37.6375 &\\
Cyg X-3 & 308.1074 & 40.9578 &\\
SAX J2103.5+4545 & 315.8990 & 45.751 &\\
Ginga 2138+56 & 324.8780 & 56.9861 &\\
Cyg X-2 & 326.1717 & 38.3217 &\\
NGC 7172 & 330.5080 & -31.8698 & T\\
4U 2206+54 & 331.9843 & 54.5185 &\\
II Peg & 358.7669 & 28.6337 &\\
\hline
\end{longtable}

\begin{table}
\tbl{Best fit equations between the BAT Crab count rate and the incident angle $\theta$.}{
\begin{tabular}{cc}
\hline
Energy band & Best fit quadratic equation\\
\hline
14-20 keV & Rate $= 0.0123169-1.15218 \times 10^{-5} \, \theta -1.4965 \times 10^{-6} \, \theta^{2}$\\
20-24 keV & Rate $= 0.0066923+6.10315 \times 10^{-6} \, \theta - 6.78433 \times 10^{-7} \, \theta^{2}$\\
24-35 keV & Rate $= 0.0105085+5.19742 \times 10^{-6} \, \theta - 6.91427 \times 10^{-7} \, \theta^{2}$\\
35-50 keV & Rate $= 0.0069828+7.88677 \times 10^{-6} \, \theta - 5.25433 \times 10^{-7} \, \theta^{2}$\\
50-75 keV & Rate $= 0.0057086+4.41883 \times 10^{-6} \, \theta - 2.67754 \times 10^{-7} \, \theta^{2}$\\
75-100 keV & Rate $= 0.0020645+8.86035 \times 10^{-7} \, \theta+ 5.79986 \times 10^{-8} \, \theta^{2}$\\
100-150 keV & Rate $= 0.0010577-6.27195 \times 10^{-6} \, \theta+ 2.51476 \times 10^{-7} \, \theta^{2}$\\
150-195 keV & Rate $= 0.0001824-3.96457 \times 10^{-6} \, \theta + 1.74441 \times 10^{-7} \, \theta^{2}$\\
\hline
\end{tabular}}\label{tbl:offaxis_corr}
\end{table}

\begin{table}
\tbl{Products of the broad-band transient monitor\\}{
\begin{tabular}{l}
\hline
Plots of daily seven channel light curves and the hardness ratio between the BAT 50-100 keV and the GSC 2-5 keV band\\
Plots of zoom-in daily light curves over the last ten days and the hardness ratio between the BAT 50-100 keV and the GSC 2-5 keV band\\
Interactive plots of daily seven channel light curves and the hardness ratio between the BAT 50-100 keV and the GSC 2-5 keV band\\
Plots of 90 minutes seven channel light curves\\
Plots of zoom-in 90 minutes seven channel light curves over the last 10 days\\
FITS light curve files of BAT four channel daily light curves\\
FITS light curve files of BAT four channel 90 minutes light curves\\
\hline
\end{tabular}}\label{tbl:bb_product_summary}
\end{table}

\end{document}